\documentclass[journal]{IEEEtran}

\usepackage[noadjust]{cite}
\usepackage{amsmath,amssymb,amsfonts,array}
\usepackage{algorithmic}
\usepackage{graphicx}
\usepackage{textcomp}
\usepackage{xcolor}
\usepackage{newunicodechar}

\begin{document}

\title{ A Quantitative Approach To The Temporal Dependency in Video Coding}

\author{Jingning~Han,~\IEEEmembership{Senior~Member,~IEEE,}
        Paul~Wilkins,
        Yaowu~Xu,~\IEEEmembership{Senior~Member,~IEEE}
        and~James~Bankoski% <-this % stops a space
\thanks{The authors are with Google LLC, Mountain View, CA 94043 USA (email: jingning@google.com).
This paper has been submitted to an IEEE publication for peer review.}}

\maketitle

\begin{abstract}
Motion compensated prediction is central to the efficiency of video compression. Its predictive coding scheme propagates the quantization distortion through the prediction chain and creates a temporal dependency. Prior research typically models the distortion propagation based on the similarity between original pixels under the assumption of high resolution quantization. Its efficacy in the low to medium bit-rate range, where the quantization step size is largely comparable to the magnitude of the residual signals, is questionable. This work proposes a quantitative approach to estimating the temporal dependency. It evaluates the rate and distortion for each coding block using the original and the reconstructed motion compensation reference blocks, respectively. Their difference effectively measures how the quantization error in the reference block impacts the coding efficiency of the current block. A recursive update process is formulated to track such dependency through a group of pictures. The proposed scheme quantifies the temporal dependency more accurately across a wide range of operating bit-rates, which translates into considerable coding performance gains over the existing contenders as demonstrated in the experiments.
\end{abstract}


\IEEEpeerreviewmaketitle

\section{Introduction}
\label{sec:intro}
\IEEEPARstart{V}{ideo} compression techniques exploit the temporal correlations in video signals, most commonly in the form of motion compensated prediction, to achieve superior coding efficiency. Such predictive coding schemes create a dependency between a coding block and its reference block. Hence, the reconstruction quality of one block can potentially influence the prediction quality and compression efficiency of blocks in subsequent frames. Measuring this temporal dependency plays a critical role in a wide range of applications in video compression, including coding block level Lagrangian multiplier and quantization parameter adaptation \cite{sdtp-icme}\nocite{mul_opt1, mbtree}-\cite{mbtree-skip}, frame level bit allocation optimization \cite{mul_opt2}-\nocite{lagrangian_opt, frame_qp, trellis_qp}\cite{waterloo}, understanding perceptual quality \cite{vmaf}, and information-theory based estimation of the rate-distortion bound \cite{info-flow}.

A source distortion temporal propagation model was developed in \cite{sdtp-icme} in the context of uni-directional motion compensated prediction \cite{mul_opt1}. Consider a pixel block $P_n$ in frame $n$ and its motion compensated reference $P_{n-1}$ in the previous frame. Following the high quantization resolution assumption, the quantization distortion in $P_n$ is modeled by
\begin{equation*}
D_n = \mathrm{e}^{-bR_n}\cdot D_{n}^{MCP},
\end{equation*}
where $b$ is related to the statistical properties of the source signal and $D_{n}^{MCP}$ represents the prediction error, which is further approximated by
\begin{equation*}
\begin{split}
D_{n}^{MCP} & = ||P_n - \hat{P}_{n-1}||^2  \\
& \approx \alpha \cdot (||\hat{P}_{n-1} - P_{n-1}||^2 + ||P_n - P_{n-1}||^2) \\
& = \alpha \cdot (D_{n-1} + D_n^{OMCP}),
\end{split}
\end{equation*}
where $\alpha$ is a constant and is empirically set to be $0.9-1$. $\hat{P}_{n-1}$ denotes the reconstruction of $P_{n-1}$. The term $D_n^{OMCP}=||P_n - P_{n-1}||^2$ represents the difference between the original pixels in $P_n$ and $P_{n-1}$, and $D_{n-1}$ is the quantization distortion in the reference block $P_{n-1}$. The relationship between the distortion terms is hence formulated as
\begin{equation}
D_n \approx \mathrm{e}^{-bR_n} \cdot \alpha \cdot (D_{n-1} + D_n^{OMCP}),
\end{equation}
where the value of $\mathrm{e}^{-bR_n}$ is obtained through a preset lookup table based on the quantization step size and the prediction error. The source distortion temporal model was further extended to handle bi-directional prediction in hierarchical pyramid coding structures \cite{mul_opt2} under certain assumptions of the similarities between statistical properties of consecutive groups of pictures \cite{waterloo}. Other simplifications include using a forward block matching scheme (i.e. conducting motion search from frame $(n-1)$ toward $n$) and assuming all inter-mode coded blocks were employed, such that the model can be built within a single pass encoding. The source distortion propagation model has a relatively low computational cost and has been successfully deployed to optimize coding decisions at both block \cite{sdtp-icme, mul_opt1} and frame levels \cite{mul_opt2, waterloo}.

A more complex approach, namely macroblock tree (MBtree), that requires a two-pass encoding is proposed in \cite{mbtree}, where the scheme accounts for hybrid intra and inter predictions, as well as the backward motion compensation (i.e. conducting motion search from frame $n$ toward $(n-1)$ as is used in the actual encoding process). The MB-tree estimates the amount of information each MB contributes to the prediction of future frames, thereby determining the distortion propagation flow. Its first pass runs with ordinary motion estimation following the frame processing order in a group of pictures. The second pass works in the reverse frame processing order over the source frames. It utilizes the motion information gathered in the first pass and propagates the dependency from future frames back to the current frame. 

For block $P_n$, the scheme estimates its intra and inter prediction errors in the form of sum of absolute transform coefficients (SATD), denoted by $S_n^{intra}$ and $S_n^{inter}$. The $S_n^{intra}$ indicates the block entropy, and the difference $(S_n^{intra} - S_n^{inter})$ reflects the amount of entropy removed by the motion compensated prediction. The correlation between $P_n$ and its reference $P_{n-1}$ is hence approximated by
\begin{equation}
\label{eq:mbtreerho}
\rho_n = \frac{S_n^{intra} - S_n^{inter}}{S_n^{intra}}.
\end{equation}
An ambient variable $C_n$ is introduced to track the total redundancy in the group of pictures being removed due to the current block $P_n$. It is propagated to its reference block $P_{n-1}$ through
\begin{equation}
\label{eq:mbtree}
C_{n-1} = \rho_n \cdot (S_n^{intra} + C_n).
\end{equation}
For a block $P_n$, it can be interpreted that it has been used as a motion compensated reference for $C_n / S_n^{intra}$ times. Therefore its quantization distortion $D_n$ will contribute an additional distortion of
\begin{equation}
\label{eq:mbdist} 
\Delta D_n =  D_n\cdot\frac{C_n}{S_n^{intra}}
\end{equation}
in the entire group of pictures. The MB-tree approach has been extended to further account for the skip mode, and to use a sigmoid function to replace the linear correlation in \eqref{eq:mbtreerho}, where the sigmoid model parameters are preset using offline training data \cite{mbtree-skip}. A similar framework has been derived using the page rank algorithm in \cite{pixelrank}.

Both the source distortion propagation model and the MB-tree are built on source signals. However, the true distortion propagation model depends on not only the temporal correlation between source pixels, but also the quantization effect. For example, in high resolution quantization setting, the quantization error is far less than the energy of the innovation term, and therefore is less likely to travel through the prediction chain. Conversely when the quantization error is comparable to or exceeds the innovation term, the distortion propagation is largely determined by the correlation between blocks in the same motion trajectory. Both the source distortion propagation model and the MB-tree adapt their propagation formulas based on the quantization step size using pre-trained parameters \cite{mul_opt1, mbtree-skip}. In our recent work \cite{vp9-tpl}, the MB-tree update model \eqref{eq:mbtree} is modified to further account for the local quantization effect:
\begin{equation}
\label{eq:vp9tpl}
C_{n-1} = \frac{D_n}{\sigma_n^2} \cdot \rho_n \cdot (S_n^{intra} + C_n),
\end{equation}
where $\sigma_n^2$ denotes the prediction error between source pixels. However, the efficacy of these modifications that use a simple linear or sigmoid function to model the quantization effect, is limited, due to high non-linearity in the quantization.

In contrast, this work proposes a framework that allows one to quantitatively evaluate the impact of the quantization distortion in the reference block on the coding efficiency of the current block. It adopts the two-pass encoding mechanism of the MB-tree but instead of running over the original frames, the first pass encoding will maintain both the original and the coded versions of each frame. A current coding block will then be predicted using both the original and the reconstructed reference blocks to obtain its rate and distortion costs. The difference between the rate and distortion costs for the two references effectively captures the impact of the quantization distortion in the reference block on the current block's coding efficiency. The second pass works in the reverse frame processing order, and recursively updates each block's impact on subsequent frames, thereby building a temporal dependency model (TPL model hereinafter).

The TPL model is implemented in the AV1 codec \cite{av1overview}. It is experimentally shown that it can closely track temporal dependency in the video coding process, which translates into considerable compression gains in the example use cases, across a wide range of test sets and quantization parameters.

\section{The TPL Framework}
\label{sec:tpl}
The proposed TPL framework consists of two processing phases, namely motion flow construction and the dependency synthesis. The first pass encoding builds the motion flow construction. It follows the frame processing order to simulate the motion compensated predictions that will be used in the final encoding pass for each group of pictures. This phase is used to estimate the compression performance difference due to the use of the original and reconstructed reference frames respectively. The second pass traverses the motion flow and propagates the block level coding statistics backward to form the dependency synthesis.

The scheme uses the quantization parameter associated with the leaf frames in a group of pictures (i.e. the B frames in a pyramid coding structure), which is given in Q mode and can be closely estimated in variable bit rate (VBR) mode, as the operating quantization parameter. It effectively measures the overall rate and distortion reductions in a group of pictures by eliminating the quantization distortion in a coding block.

\subsection{Motion Flow Construction}
\label{sec:mf_construction}
Consider a coding block $P_n$ in frame $n$. The motion flow construction will check the inter prediction modes, including both single and bi-directional predictions, over all the reference frames. In AV1, there are a maximum of 7 available reference frames \cite{av1overview}. To keep the computational complexity under check, the scheme uses a fixed $16\times16$ processing block size. The motion vectors and the associated reference frame indexes that minimize the rate-distortion cost will be stored per coding block. The inter mode search is conducted over both the original and the reconstructed reference frames, respectively. The resulting rate and distortion costs for the original and reconstructed reference frames, denoted by $(R_{src}(n), D_{src}(n))$ and $(R_{rec}(n), D_{rec}(n))$, are stored on a block basis. For single reference frame mode, the difference
\begin{eqnarray}
\label{eq:rddiff}
\delta D(n) &=& D_{rec}(n) - D_{src}(n) \\ 
\delta R(n) &=& R_{rec}(n) - R_{src}(n)
\end{eqnarray}
captures the coding statistics change due to the quantization distortion in the reference. 

For bi-directional prediction modes, where two reference frames are involved, we evaluate each individual reference frame's impact separately by comparing the coding statistics using its original and reconstructed forms, but only the reconstructed reference for the other frame. Let $R_{src, rec}(n)$ and $D_{src, rec}(n)$ denote the rate and distortion costs using the original first reference frame and the reconstructed second reference frame. The coding performance change in $P_n$ due to the quantization distortion in the first reference frame is hence measured by
\begin{eqnarray}
\label{eq:cmpdiff}
\delta D(n) &=& D_{rec}(n) - D_{src, rec}(n) \\ 
\delta R(n) &=& R_{rec}(n) - R_{src, rec}(n),
\end{eqnarray}
where $(R_{rec}(n), D_{rec}(n))$ are the rate and distortion costs using the reconstructed reference frames on both sides, following the above convention. Similarly, the impact of the distortion in the second reference frame is evaluated by changing the condition of the second reference frame, in the context of a reconstructed first reference frame:
\begin{eqnarray}
\label{eq:cmpdiff2}
\delta D(n) &=& D_{rec}(n) - D_{rec, src}(n) \\ 
\delta R(n) &=& R_{rec}(n) - R_{rec, src}(n),
\end{eqnarray}
where $R_{src, rec}(n)$ and $D_{src, rec}(n)$ denote the rate and distortion cost using the reconstructed first reference frame and the original second reference frame. In the rare cases where either $\delta D(n) < 0$ or $\delta R(n) < 0$, we lower bound it at 0.

Intra prediction modes are evaluated using only the reconstructed spatial neighboring pixels to form the prediction. If an intra mode provides lower rate-distortion cost than the above inter modes using the source reference frames, the scheme will mark it for the block and reset $\delta D(n)$ and $\delta R(n)$ to 0. Unlike the MB-tree and its variants that compare the intra and inter prediction results to form the distortion propagation factor \eqref{eq:mbtreerho}, the TPL framework compares them for a binary decision. If an intra mode predicts the block better, the temporal dependency on its motion compensated reference will be completely cut off. Otherwise, the intra prediction results will not affect the temporal dependency calculation.

\subsection{Dependency Synthesis}
Having established the relationship between the current block and its reference block, we now propagate such information backward through the motion compensated prediction chain to build the temporal dependency. Let $\Delta R(n)$ and $\Delta D(n)$ denote the \emph{total} additional  rate and distortion costs in the subsequent group of pictures caused by the quantization distortion, $D_{rec}(n)$, in the current block $P_n$. We derive the corresponding terms for the reference block, $P_{n-1}$.

We first consider the additional distortion in the group of pictures caused by the quantization distortion in $P_{n-1}$, i.e. $\Delta D(n-1)$. The immediate impact of the distortion in $P_{n-1}$ is on $P_{n}$, which is captured by $\delta D(n)$ as discussed above. The $\Delta D(n)$ is associated with the actual distortion in $P_n$, that is the quantization distortion using the reconstructed reference, $D_{rec}(n)$, of which $\delta D(n)$ contributes to the distortion in $P_{n-1}$. A linear model is used here to estimate the amount of distortion in $\Delta D(n)$ that contributes to the distortion in $P_{n-1}$, i.e. $\frac{\delta D(n)}{D_{rec}(n)} \Delta D(n)$. Therefore the update procedure for $\Delta D(n-1)$ is formulated as
\begin{equation}
\label{eq:dist-update}
\Delta D(n-1) = \delta D(n) + \frac{\delta D(n)}{D_{rec}(n)} \cdot \Delta D(n).
\end{equation}

Next we consider the additional rate cost in the group of pictures incurred by the quantization distortion in $P_{n-1}$, denoted by $\Delta R(n-1)$. For the same quantizer, the rate cost is logarithmic to the prediction errors \cite{comp_book}. Hence the additional rate cost due to $D_{rec}(n)$ can be approximated by
\begin{equation}
\label{eq:rate-form}
\Delta R(n) \approx \frac{1}{2}\cdot log_2 \frac{D_{rec}(n) + \sigma_{n+1}^2}{\sigma_{n+1}^2},
\end{equation}
where $\sigma_{n+1}^2$ is the prediction error in the next block in the motion trajectory using the original $P_n$ as the reference. The term $\sigma_{n+1}^2$ is not directly measurable in the update process, since we only track the temporal dependency model for on-grid blocks in each frame, whereas the reference block for subsequent frames is often off-grid. Also note that here we assume the quantization noise $D_{rec}$ and the innovation term $\sigma_{n+1}^2$ are uncorrelated.

As measured in Sec \ref{sec:mf_construction} the distortion in $P_{n-1}$ introduces an additional rate cost $\delta R(n)$ in $P_n$. It also changes the quantization distortion in $P_n$ from $D_{src}(n)$ to $D_{rec}(n)$. Following \eqref{eq:rate-form}, this translates into an additional rate cost for the remaining blocks in the motion compensated prediction chain:
\begin{equation}
\label{eq:rate-term}
\begin{split}
\Delta R'(n) &= \frac{1}{2}\cdot log_2\frac{D_{rec}(n) + \sigma_{n+1}^2}{D_{src}(n) + \sigma_{n+1}^2} \\
&= \frac{1}{2}\cdot log_2\frac{\frac{D_{rec}(n)}{\sigma_{n+1}^2} + 1}{\frac{D_{rec}(n)}{\sigma_{n+1}^2} \cdot \frac{D_{src}(n)}{D_{rec}(n)} + 1}.
\end{split}
\end{equation}
From \eqref{eq:rate-form}, we have
\begin{equation}
\label{eq:inno-term}
\frac{D_{rec}(n)}{\sigma_{n+1}^2} \approx 2^{2\Delta R(n)} - 1.
\end{equation}
Plug \eqref{eq:inno-term} in \eqref{eq:rate-term}, we have
\begin{equation}
\Delta R'(n) = log_2 \frac{2^{2\Delta R(n)}}{\frac{D_{src}(n)}{D_{rec}(n)} \cdot 2^{2\Delta R(n)} + (1 - \frac{D_{src}(n)}{D_{rec}(n)})}.
\end{equation}
Therefore the total additional rate cost in the group of pictures that contributes to the distortion in $P_{n-1}$ is
\begin{equation}
\label{eq:rate-update}
\begin{split}
\Delta R(n-1) &= \delta R(n) + \Delta R'(n) \\
&= \delta R(n) + log_2 \frac{2^{2\Delta R(n)}}{\frac{D_{src}(n)}{D_{rec}(n)} \cdot 2^{2\Delta R(n)} + 1 - \frac{D_{src}(n)}{D_{rec}(n)}}.
\end{split}
\end{equation}

\begin{figure}[h]
\includegraphics[width=0.47\textwidth]{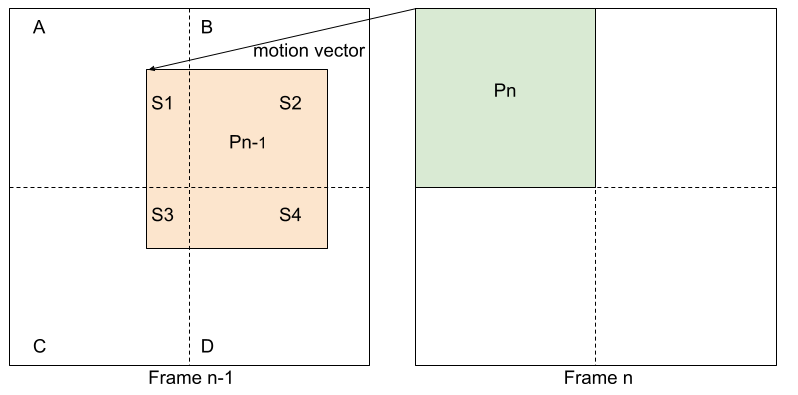}
\centering
\caption{An illustration of a reference block $P_{n-1}$ that sits off-grid. The TPL scheme distributes the calculated $\Delta R(n-1)$ and $\Delta D(n-1)$ among the on-grid blocks that it overlaps with.}
\label{fig:offgrid}
\end{figure}

In a group of pictures consisting of $M$ frames, all $\Delta R(M)$ and $\Delta D(M)$ are initialized to be 0. The TPL scheme recursively updates $\Delta R(n)$ and $\Delta D(n)$ from $M$-th frame to the first frame on a block basis. It tracks and updates the dependency model only for on-grid blocks. When a reference block $P_{n-1}$ sits off-grid, the calculated $\Delta R(n-1)$ and $\Delta D(n-1)$ will be linearly distributed among the overlapping on-grid blocks. For example, in Figure \ref{fig:offgrid}, block A will get $\frac{S_1}{S_1 + S_2 + S_3 + S_4}\Delta R(n-1)$ and  $\frac{S_1}{S_1 + S_2 + S_3 + S_4}\Delta D(n-1)$. This information distribution approach is similar to that used in \cite{mbtree} and \cite{delayed_decoding} as well.

For a given block $P_n$, the total distortion that it contributes to the group of pictures is $D_{rec}(n) \cdot (1 + \frac{\Delta D(n)}{D_{rec}(n)})$, where the term 
\begin{equation}
\label{eq:tpl-dist-prop}
\beta_n = \frac{\Delta D(n)}{D_{rec}(n)}
\end{equation}
reflects the temporal distortion dependency on $P_n$. Similarly the total rate cost for the group of pictures contributed by coding $P_n$ is $R_{rec}(n) + \Delta R(n)$.

\section{Lagrangian Multiplier Scaling}
\label{sec:mulscl}
As noted earlier, the temporal dependency model has a wide range of applications in video coding. We consider the Lagrangian multiplier scaling at the largest coding block (LCB) level here to demonstrate that the improved temporal dependency model provided by the TPL scheme translates into compression gains.

Let $D_{rec}(n, k)$ denote the quantization distortion in block $k$ in frame $n$. Similarly let $\Delta R(n, k)$ and $\Delta D(n, k)$ denote the total additional rate and distortion costs for the group of pictures caused by $D_{rec}(n, k)$. The Lagrangian multiplier associated with the quantization parameter used in the TPL scheme, i.e. the leaf frames' quantization parameter (see Section \ref{sec:tpl}), is denoted by $\lambda_{tpl}$. The base Lagrangian multiplier associated with the quantization parameter of frame $n$ used in the final encoding process is $\lambda_n$.

The additional rate-distortion cost for the group of pictures caused by $D_{rec}(n, k)$ is $\Delta D(n, k) + \lambda_{tpl} \Delta R(n, k)$. For LCB $m$, we define its distortion propagation factor $\alpha_m$ as
\begin{equation}
\alpha_m = \frac{\sum_{k\in LCB_m}\Delta D(n, k) + \lambda_{tpl} \Delta R(n, k)}{\sum_{k\in LCB_m}D_{rec}(n, k)},
\end{equation}
where the summations in both the numerator and denominator are over all operating blocks in LCB $m$. Note that $\sum_{k\in LCB_m}D_{rec}(n, k) \cdot (1 + \alpha_m)$ effectively captures the overall rate-distortion cost for the group of pictures that attributes to the distortion in LCB $m$. Similarly one can calculate the distortion propagation factor $\alpha_{fr}$ for frame $n$:
\begin{equation}
\alpha_{fr} = \frac{\sum_{k}\Delta D(n, k) + \lambda_{tpl} \Delta R(n, k)}{\sum_{k}D_{rec}(n, k)},
\end{equation}
where the summations go over all the operating blocks in frame $n$. We use $\alpha_{fr}$ to normalize $\alpha_m$ and adapt the Lagrangian multiplier at LCB $m$ as \cite{rdopt}
\begin{equation}
\label{eq:mulscl}
\lambda_{m} = \lambda_n \cdot \frac{1 + \alpha_{fr}}{1 + \alpha_m}.
\end{equation}

A similar approach has been used in the context of MB-tree \cite{mbtree, vp9-tpl}. Using \eqref{eq:mbdist}, one can derive the distortion propagation factor for LCB $m$ using the MB-tree:
\begin{equation}
\label{eq:mbtreedist}
\alpha_m^{MB} = \frac{\sum_{k\in LCB_m} C_{n,k}}{\sum_{k\in LCB_m}S_{n, k}^{intra}},
\end{equation}
where $S_{n, k}^{intra}$ represents the SATD costs of the intra prediction mode for the $k$-th block in frame $n$, and $C_{n,k}$ denotes the additional SATD reductions due to using this block for motion compensation. The frame level distortion propagation factor is
\begin{equation}
\label{eq:mbtreedistfrm}
\alpha_{fr}^{MB} = \frac{\sum_{k} C_{n,k}}{\sum_{k}S_{n, k}^{intra}},
\end{equation}
where the summations go through all the operating blocks in frame $n$. Following \eqref{eq:mulscl}, the Lagrangian multiplier for the LCB $m$ is obtained by
\begin{equation}
\lambda_{m}^{MB} = \lambda_n \cdot \frac{1 + \alpha_{fr}^{MB}}{1 + \alpha_m^{MB}}.
\end{equation}

We evaluate the compression performance of Lagrangian multiplier scaling based on the TPL and MB-tree schemes, respectively, and experimentally demonstrate that the improved temporal dependency model can translate into considerable compression gains in this setting.

\section{Experimental Results}
The experiments were conducted in the libaom AV1 codec \cite{libaom}. The baseline used the high latency (2-pass encoding) constant quantization parameter (QP) mode, where the leaf frames' QP was provided as an input to the encoder. The QP offsets for other frames in lower layers of a pyramid coding structure were determined based on the frame correlations calculated in a first-pass encoding process. The encoder settings were
\begin{verbatim}
./aomenc input_file.y4m
-o output_bitstream.webm
--end-usage=q --cq-level=LEAF-QP
--cpu-used=1
--passes=2
\end{verbatim}
Here --cpu-used=1 indicates a high complexity and high compression efficiency coding mode.

We first compare the model accuracy between the TPL and the MB-tree variants, as they both require a 2-pass encoding and use high computational complexity to achieve accurate modeling. We then demonstrate that the improved model accuracy can translate into compression gains in video coding.

\subsection{Model Accuracy}
\label{sec:modeleval}
A key element of the temporal dependency model is the distortion propagation factor \eqref{eq:tpl-dist-prop} that evaluates the total additional distortion for the group of pictures caused by the quantization distortion in a current block.

To observe the reference data, we modified the baseline to use a fixed group of picture length of 16 frames each. We considered the second group of pictures, i.e. coding frame 17 - 32. The first run used the default high latency constant QP mode and we tracked the total distortion for this group of pictures as $D_1$, and the distortion $d_1$ for the frame at the lowest layer in the pyramid coding structure, i.e. frame 32 which served as the long term reference frame for this group of pictures. We then slightly adjusted the QP for frame 32 and left the remaining frames' settings unchanged. We re-ran the coding process to gather the distortion $D_2$ for this group of pictures and the distortion $d_2$ for frame 32.

The distortion perturbation in the long term reference frame
\begin{equation}
\delta d = d_2 - d_1
\end{equation}
caused the change of the overall distortion for the entire group of pictures
\begin{equation}
\Delta D = D_2 - D_1.
\end{equation}
Therefore the distortion propagation factor at this set of operating QPs for the group of pictures is calculated by
\begin{equation}
\beta_{obs} = \frac{\Delta D}{\delta d} - 1
\end{equation}
where the minus 1 on the right hand side reflects the fact that $\Delta D$ includes $\delta d$. We varied the leaf frames' QPs to observe the distortion propagation factors at different operating points, and used these observations as ground truth.

The TPL framework estimates the distortion propagation factor at coding block level using \eqref{eq:tpl-dist-prop}. This can be extended to estimate the distortion propagation at frame level
\begin{equation}
\label{eq:tpl-dist-frame}
\beta_{TPL} = \frac{\sum_{k \in frame\ n}\Delta D(k, n)}{\sum_{k \in frame\ n}D_{rec}(k, n)}.
\end{equation}
In this context, $n=32$.

The MB-tree estimates the distortion propagation factor at coding block level though \eqref{eq:mbdist}. Similarly the frame level distortion propagation can be summarized by
\begin{equation}
\label{eq:mb-dist-frame}
\beta_{MB-tree} = \frac{\sum_{k \in frame\ n}C_{n, k}}{\sum_{k \in frame\ n}S_{n, k}^{intra}}.
\end{equation}
A variant of MB-tree \cite{vp9-tpl} that further accounts for the quantization effect in the backward update process \eqref{eq:vp9tpl} is also included in the model accuracy evaluation, and is referred to as MB-tree-Quant. It shares the same frame level distortion propagation estimation as MB-tree \eqref{eq:mb-dist-frame}.

\begin{figure}[h]
\includegraphics[width=0.45\textwidth]{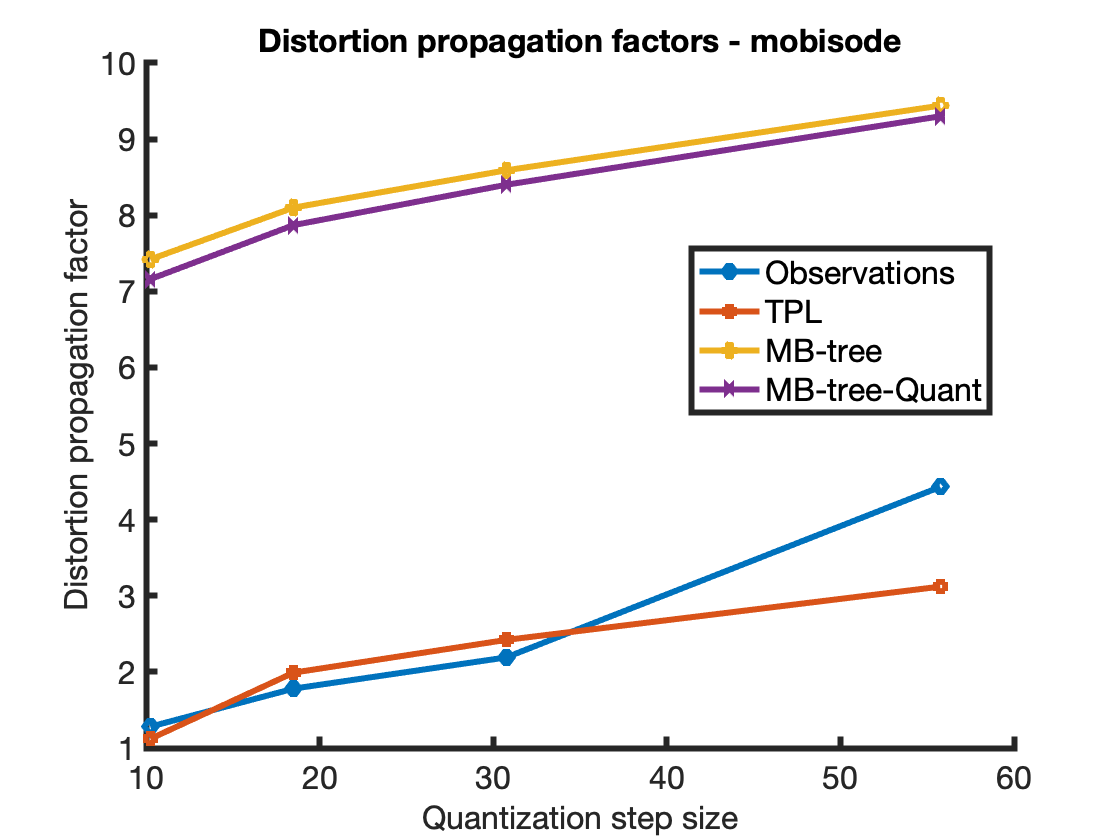}
\centering
\caption{The comparison of the distortion propagation factor estimates using TPL, MB-tree, and MB-tree-Quant. The ground truth data is marked as Observations. The test clip is $mobisode$ at 480p resolution.}
\label{fig:dist-mobi}
\end{figure}

\begin{figure}[h]
\includegraphics[width=0.45\textwidth]{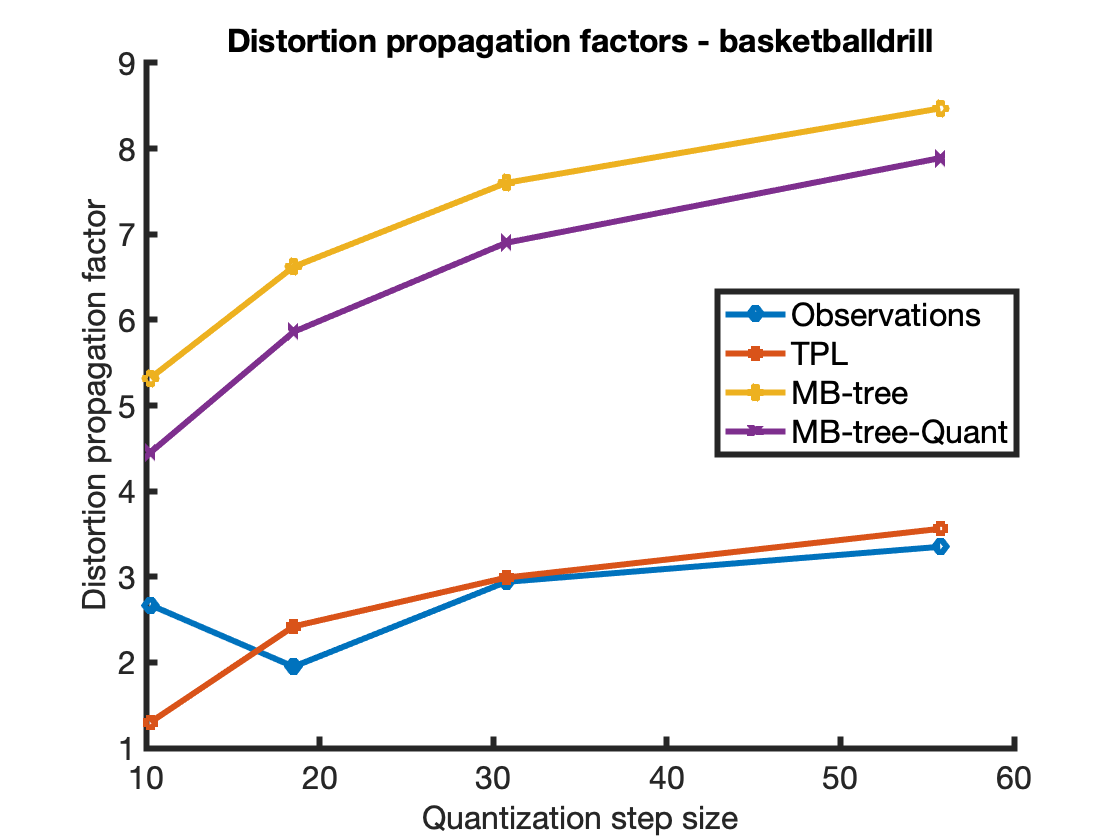}
\centering
\caption{The comparison of the distortion propagation factor estimates using TPL, MB-tree, and MB-tree-Quant. The ground truth data is marked as Observations. The test clip is $BasketBallDrill$ at 480p resolution.}
\label{fig:dist-basketball}
\end{figure}

\begin{figure}[h]
\includegraphics[width=0.45\textwidth]{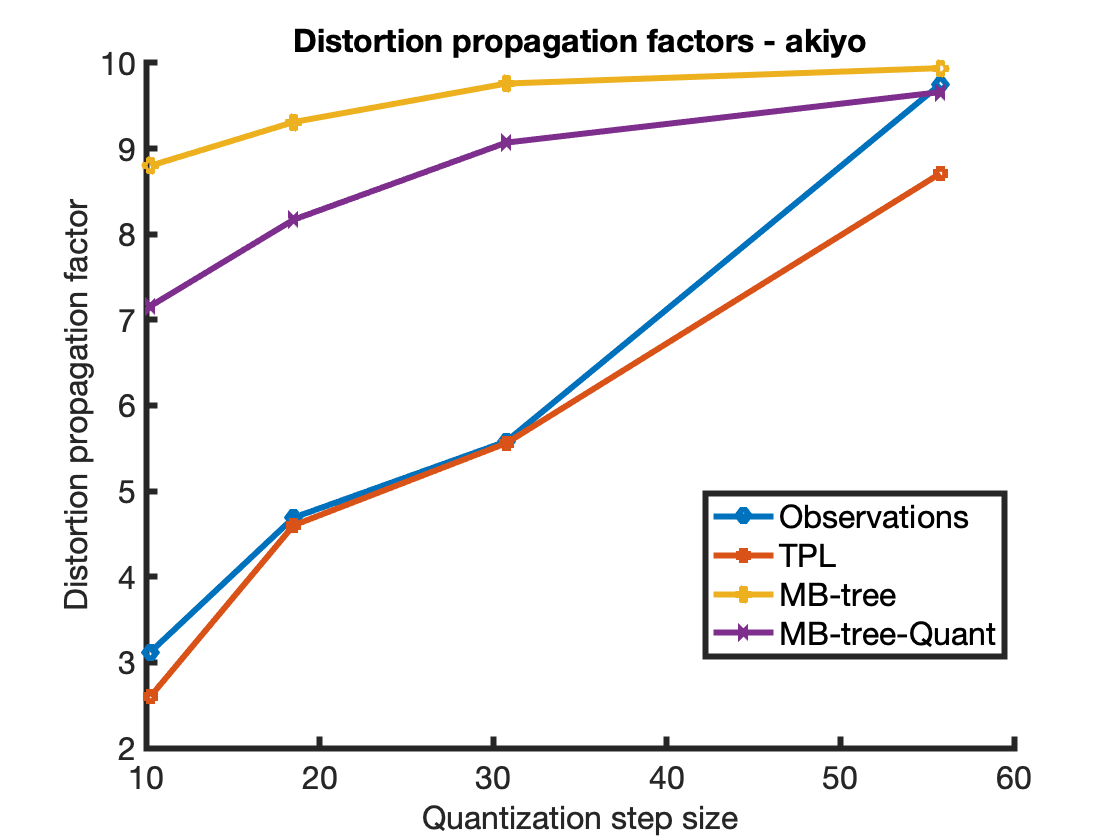}
\centering
\caption{The comparison of the distortion propagation factor estimates using TPL, MB-tree, and MB-tree-Quant. The ground truth data is marked as Observations. The test clip is $akiyo$ at CIF resolution.}
\label{fig:dist-akiyo}
\end{figure}

The above distortion propagation factor estimates are compared against the ground truth in Figure \ref{fig:dist-mobi}-\ref{fig:dist-akiyo}. For each sample clip, we test 4 leaf frame QPs whose quantization step sizes range from 10 to 56, which represents a wide spectrum of fidelity. The horizontal axis is the quantization step size applied to the transform coefficients. The vertical axis shows the distortion propagation factor. The ground truth data is labeled as ``Observations''. The estimates provided by TPL \eqref{eq:tpl-dist-frame}, MB-tree \eqref{eq:mb-dist-frame}, and MB-tree-Quant \eqref{eq:vp9tpl} and \eqref{eq:mb-dist-frame} are denoted by ``TPL'', ``MB-tree'', and ``MB-tree-Quant'' respectively.

It is demonstrated that the MB-tree tends to over estimate the distortion propagation factor, since it mainly operates on source pixels. The MB-tree-Quant consistently improves the estimation accuracy by further accounting for the quantization effect \eqref{eq:vp9tpl} derived under the high resolution quantization assumption \cite{vp9-tpl}, but is still distant from the ground truth. The TPL framework instead uses a quantitative approach to measure the quantization effect \eqref{eq:dist-update}, which allows it to track the ground truth data closely across the operating QPs and variations of video signal statistics.

\subsection{Compression Efficiency Gains}
Next we used the Lagrangian multiplier scaling scheme discussed in Sec \ref{sec:mulscl} as an example to demonstrate that the more accurate estimate of the temporal dependency can translate into actual compression efficiency gains. Here we used the same 2-pass high latency constant QP mode as mentioned above. To make the baseline achieve better compression efficiency, the codec employed an adaptive group of picture length approach \cite{goplen}. The test clips covered $480p$ to $1080p$ resolutions. The operating points were selected to make the reconstruction quality span a range of $~35dB$ to $~45dB$ in PSNR for each test clip.

As shown in Sec \ref{sec:modeleval}, the MB-tree-Quant provided small but fairly consistent model accuracy improvements over the MB-tree. Hence we implemented the Lagrangian multiplier scaling using the MB-tree-Quant as the reference. The Lagrangian multiplier scaling based on the proposed TPL framework was implemented on top of the same baseline. The compression efficiency gains of the MB-tree-Quant reference and the TPL over the baseline were shown in Table \ref{tab:midres} and \ref{tab:hdres}. It was shown that the TPL based approach achieved more coding gains in the Lagrangian multiplier scaling scheme that accounted for the temporal dependency in the rate-distortion optimization \cite{rdopt} framework.

We also included a complexity evaluation for the proposed TPL framework. We used a single thread to encode both the baseline and its extension using the TPL based Lagrangian multiplier scaling. The CPU instruction counts were compared to estimate the additional encoder complexity due to TPL and are shown as a percentage in Table \ref{tab:midres} and \ref{tab:hdres}. Note that there are many implementation designs that could reduce the TPL complexity, which are beyond the scope of this paper and are not included in the complexity evaluation here.

\begin{table*}[htb]
\caption{The compression efficiency gains of the Lagrangian multiplier scaling using the MB-tree-Quant and the proposed TPL over the baseline. The performance is evaluated in BD-rate reduction using PSNR and SSIM as quality metrics. A negative number means better compression efficiency. The test set consists of $480p$ resolution videos.}
\begin{center}
\begin{tabular}{|c|c|c|c|c|c|}
\hline
  & \multicolumn{2}{c|}{MB-tree} & \multicolumn{3}{c|}{TPL} \\
\hline
 &  PSNR & SSIM &  PSNR & SSIM & encoder complexity change \\
\hline
BQMall\_832x480\_60	 & -1.039 &	-2.707 &	-2.386 &	-5.664 & 4.19\%  \\
BalloonFestival\_854x480\_24fps &	1.332 &	3.049 &	-1.037 &	-8.113 & 2.66\% \\
BasketballDrillText\_832x480\_50 &	-2.946 &	-2.358 &	-7.392 &	-16.189 & 0.29\% \\
BasketballDrill\_832x480\_50 & 	-3.792 &	-3.793 &	-8.637 &	-17.71 & 0.99\% \\
Campfire\_854x480\_30fps & 	0.823 &	-3.288 & 	-3.564 & 	-13.135 & 7.95\% \\
CatRobot\_854x480\_60fps &	-1.187 &	0.103 &	-2.276 & 	-3.601 & 2.16\% \\
DaylightRoad2\_854x480\_60fps &	-1.196 & 	-0.734 & 	-2.196 &	-2.182 & 1.89\% \\
Drums\_854x480\_100fps &	0.616 &	0.303 & 	-0.544 &	-4.219 & 1.89\% \\
Keiba\_832x480\_30 &	-2.846 &	-4.769 &	-4.571 &	-6.626 & 9.09\% \\
Market3Clip4000r2\_854x480\_50fps &	0.285 &	-0.059 &	-3.35 &	-11.071 & 1.90\%  \\
Mobisode2\_832x480\_30 & 	0.9 &	 0.424 &	1.204 &	-0.166  &  11.08\% \\
Netflix\_Narrator\_850x480\_60fps & 	-0.355 &	-0.64 & 	-1.741 &	-4.547 & 4.22\% \\
PartyScene\_832x480\_50 &	-1.811 &	-1.3 &	-3.244 &	-7.158 &   0.45\% \\
RaceHorses\_832x480\_30 &	-0.859 &	-2.905 &	-2.903 &	-4.344 & 6.90\% \\
ShowGirl2TeaserClip4000\_854x480\_24fps &	0.479 &	1.355 &	-1.824 &	-2.125 & 7.35\% \\
Tango\_854x480\_60fps &	- 0.056 & 	-0.781 &	-0.67 & 	-2.583 & 4.48\% \\
ToddlerFountain\_854x480\_60fps &	1.413 &	-1.155 &	-1.974 &	-3.998 & 2.34\% \\
TrafficFlow\_854x480\_30fps & 	-0.761 &	-1.475 &	-3.611 &	-10.146 & 3.22\% \\
aspen\_480p & 	0.105 &	0.223 &	-2.493 &	-4.225 & 3.22\% \\
blue\_sky\_480p25 &	 -3.378 & 	0.819 &	-4.552 & 	-5.758 & 4.33\% \\
city\_4cif\_30fps &	-0.417 &	-1.12 &	-1.885 &	-3.879 &  6.41\% \\
controlled\_burn\_480p &	0.469 &	2.034 &	-0.36 & 	-0.94 & 3.60\% \\
crew\_4cif\_30fps &	-1.221 &	-3.557 &	-3.477 &	-5.029 & 9.66\% \\
crowd\_run\_480p &	-1.307 &	-2.328 &	-2.742 &	-6.958 & 1.00\% \\
ducks\_take\_off\_480p &	-0.54 &	-2.76 &	-2.711 &	-3.766 & 2.30\%   \\
harbour\_4cif\_30fps &	1.065 &	-1.545 &	-0.089 &	-2.485 &  3.92\% \\
ice\_4cif\_30fps & 	-2.216 &	-4.135 &	-3.172 &	-8.718 & 10.40\% \\
into\_tree\_480p &	-0.337 &	2.311 &	-1.767 &	-3.535 & 3.74\% \\
netflix\_aerial &	0.042 & 	0.596 &	-2.735 &	-4.072 & 0.84\% \\
netflix\_foodmarket &	-1.818 &	0.032 &	-2.944 &	-4.45 & 5.44\% \\
netflix\_ritualdance & 	-1.415 &	-1.383 &	-3.7 &	-7.616 & 7.14\% \\
netflix\_rollercoaster &	0.158 &	0.529 &	-2.201 &	-5.474 & 5.83\% \\
netflix\_squareandtimelapse &	-0.199 &	0.599 &	-1.9 &	-4.16 & 1.80\% \\
netflix\_tunnelflag &	-0.119 &	0.106 &	-2.202 &	-2.599 & 3.77\% \\
old\_town\_cross\_480p &	0.256 &	1.59 & 	-0.173 &	-0.196 & 3.33\% \\
park\_joy\_480p &	-1.763 &	1.585 &	-4.28 &	-5.858 & 1.32\% \\
red\_kayak\_480p &	0.722 &	-0.275 &	-1.079 &	-1.531 & 12.81\% \\
rush\_field\_cuts\_480p & 	-0.893 &	1.712 &	-2.017 &	-2.946 &  2.10\% \\
shields\_640x360\_60 &	1.333 &	1.508 &	0.872 &	-1.005 & 2.54\% \\
sintel\_shot\_854x480\_24fps &	 -0.07 &	-2.781 &	-2.684 &	-5.048 & 7.53\% \\
soccer\_4cif\_30fps &	-1.32 &	-3.146 &	-3.738 &	-5.832 & 11.85\% \\
speed\_bag\_480p &	 2.387 &	1.38 &	0.948 &	-0.857 & 10.92\% \\
station2\_480p25 &	2.236 &	2.789 &	1.338 &	0.203 &   5.82\% \\
tears\_of\_steel1\_480p & 	-0.322 &	-0.726 &	-2.433 &	-4.945 & 5.58\% \\
touchdown\_pass\_480p &	-0.085 &	-0.023 &	-3.993 &	-10.095  & 8.19\% \\
west\_wind\_easy\_480p &	-1.664 &	-2.328 &	-4.675 &	-3.415  & 6.43\% \\
\hline
Overall & -0.46 &	-0.62 &	-2.37 &	-5.07 & 4.09\%\\
\hline
\end{tabular}
\end{center}
\label{tab:midres}
\end{table*}%

\begin{table*}[htb]
\caption{The compression efficiency gains of the Lagrangian multiplier scaling using the MB-tree-Quant and the proposed TPL over the baseline. The performance is evaluated in BD-rate reduction using PSNR and SSIM as quality metrics. A negative number means better compression efficiency. The test set consists of $720p$ and $1080p$ resolution videos.}
\begin{center}
\begin{tabular}{|c|c|c|c|c|c|}
\hline
  & \multicolumn{2}{c|}{MB-tree} & \multicolumn{3}{c|}{TPL} \\
\hline
 &  PSNR & SSIM &  PSNR & SSIM & encoder complexity change \\
\hline
BalloonFestival\_1280x720\_24fps & 	-1.898 &	-6.366 &	-1.742 &	-8.946 & 3.25\%\\
CSGO\_1080p60 &	-2.701 &	-1.701 &	-3.098 &	-4.767 & 6.84\% \\
Campfire\_1280x720\_30fps &	-1.504 &	-5.078 &	-4.003 & 	-12.786 & 2.19\% \\
CatRobot\_1280x720\_60fps &	-3.744 &	-3.445 &	-3.538 &	-4.248  & 5.66\%\\
DaylightRoad2\_1280x720\_60fps &	-2.962 &	-2.856 &	-3.069 &	-2.924 & 3.35\% \\
Drums\_1280x720\_100fps & 	-3.103 &	-6.248 &	-3.252 &	-8.064 & 5.66\%\\
Market3Clip4000r2\_1280x720\_50fps &	-4.489 &	-8.667 &	-5.65 &	-13.534 & 5.52\% \\
Netflix\_Aerial\_2048x1080\_60fps & 	-2.055 &	-1.078 &	-2.359 &	-2.029 & 6.34\%\\
Netflix\_Boat\_2048x1080\_60fps &	-6.389 &	-9.992 &	-6.301 &	-11.547 & 4.64\% \\
Netflix\_Crosswalk\_2048x1080\_60fps & 	-1.417 &	-2.328 &	-1.83 &	-3.114 & 9.53\% \\
Netflix\_Dancers\_1280x720\_60fps &	-1.818 &	-5.422 &	-0.822 &	-5.006 & 7.26\%\\
Netflix\_DrivingPOV\_2048x1080\_60fps &	-2.181 &	-2.985 &	-2.059 &	-3.48  & 3.77\% \\
Netflix\_FoodMarket2\_1280x720\_60fps &	-1.288 &	-1.86	 &  -1.309 &	-3.412 & 7.96\%  \\
Netflix\_FoodMarket\_2048x1080\_60fps &	-2.758 &	-4.43 & 	-3.283 &	-3.988 & 10.00\% \\
Netflix\_PierSeaside\_2048x1080\_60fps &	-9.031 &	-4.176 &	-7.495 &	-1.36 & 8.05\% \\
Netflix\_SquareAndTimelapse\_2048x1080\_60fps & 	-2.545 &	-4.382 &	-2.297 &	-5.539 & 7.50\% \\
Netflix\_TunnelFlag\_2048x1080\_60fps & 	-4.667 &	-5.787 &	-5.182 &	-6.03 & 6.57\% \\
RollerCoaster\_1280x720\_60fps &	-3.14 & 	-4.901 &	-3.153 &	-5.778 & 7.45\% \\
ShowGirl2TeaserClip4000\_1280x720\_24fps & 	-2.912 &	-3.461 &	-3.31 & 	-4.453 & 5.69\% \\
Tango\_1280x720\_60fps &	-1.013 &	-1.775 &	-1.003 &	-1.864 & 7.11\% \\
ToddlerFountain\_1280x720\_60fps &	-1.11 & 	-3.79 &	-1.923 &	-4.246 & 0.51\% \\
TrafficFlow\_1280x720\_30fps & 	-5.994 &	-9.483 &	-6.313 &	-11.028 & 6.06\% \\
aspen\_1080p &	-3.319 &	-7.174 &	-4.01 & 	-7.324  & 6.59\% \\
basketballdrive\_1080p50 & 	-3.937 &	-7.769 &	-4.407 &	-7.673 & 7.55\% \\
cactus\_1080p50 &	-6.403 &	-10.121 &	-6.526 &	-10.35 & 3.22\% \\
city\_720p30 &	-0.777 &	-2.392 &	-0.797 &	-2.303 & 4.82\% \\
controlled\_burn\_1080p &	-4.078 &	-3.13 &	-4.406 &	-3.563 & 5.98\% \\
crew\_720p30 &	-1.175 &	-4.776 &	-2.818 &	-7.958 & 4.18\% \\
crowd\_run\_1080p50 &	-3.346 &	-4.168 &	-3.397 &	-4.634 &  5.02\% \\
dinner\_1080p30 &	-1.46 &	-2.686 &	-2.118 &	-3.781 & 7.09\% \\
ducks\_take\_off\_1080p50 &	-0.964 &	-2.388 &	-1.981 &	-2.532 & 3.68\% \\
factory\_1080p30 &	-7.163 &	-7.442 &	-7.548  &	-8.923 & 9.87\% \\
in\_to\_tree\_1080p50 &	-1.035 &	-1.271 &	-3.485 &	-3.474 & 0.53\% \\
johnny\_720p60 &	-3.843 &	-2.24 &	-4.504 &	-3.784 & 5.76\% \\
kristenandsara\_720p60 &	 -4.369 &	-2.276 &	-4.77 &	-3.782  & 6.37\% \\
night\_720p30 &	-4.626 &	-6.506 &	-4.774 &	-8.279 & 3.25\% \\
old\_town\_cross\_720p50 &	-0.905 &	-0.292 &	-0.747 &	0.306 & 3.91\% \\
parkjoy\_1080p50 &	-3.565 &	-4.389 &	-3.672 &	-4.387 & 3.91\% \\
ped\_1080p25 &	-5.373 &	-8.188 &	-6.898 &	-11.78  & 8.00\% \\
red\_kayak\_1080p &	0.898 &	0.737 &	0.713 &	0.712 & 3.53\% \\
riverbed\_1080p25 &	 -0.401 &	-0.788 &	-0.946 &	-1.058 & 3.94\% \\
rush\_field\_cuts\_1080p &	-1.024 &	-4.728 &	-1.218 &	-4.313 & 3.58\% \\
rush\_hour\_1080p25 & 	-0.711 &	-2.819 &	-0.949 &	-2.44 & 4.00\% \\
shields\_720p50 &	-1.491 &	-4.207 &	-1.395 &	-3.584 & 5.08\% \\
station2\_1080p25 & 	-0.187 &	-1.645 &	0.097 &	-0.23  & 8.10\% \\
sunflower\_720p25 &	-2.654 &	-4.839 &	-3.846 &	-7.101 & 8.65\% \\
tennis\_1080p24 &	0.725 &	0.399 &	-0.149 &	-0.815 & 9.88\% \\
touchdown\_pass\_1080p &	-3.454 &	-6.477 &	-4.365 &	-9.078 & 6.62\% \\
tractor\_1080p25 &	-2.659 &	-3.807 &	-3.265 &	-4.121 & 7.68\% \\
vidyo4\_720p60 &	-4.912 &	-4.892 &	-6.148 &	-6.811 & 6.66\% \\
\hline
Overall &	-2.819 &	-4.21 & 	-3.226 &	-5.224 & 5.76\% \\
\hline
\end{tabular}
\end{center}
\label{tab:hdres}
\end{table*}%

The source code is available at the open source repository \cite{libaom}. Please refer to $tpl\_model.c$ file for the related software implementation.

\section{Conclusions}
A quantitative approach to temporal dependency estimation in video coding is proposed. It overcomes a longstanding difficulty in properly accounting the quantization effect for temporal dependency modeling due to its non-linearity. Implemented in the libaom AV1 codec, it is experimentally demonstrated that the proposed framework provides a more accurate temporal dependency estimate across a wide range of operating points and variations of video signal statistics. This improvement is further demonstrated to translate into compression efficiency gains in a sample application.

\bibliographystyle{IEEEtran}
\bibliography{tpl}

\end{document}